# Human Factors in Biocybersecurity Wargames

Lucas Potter [1], Xavier-Lewis Palmer[1]

[1] 1 Biomedical Engineering Institute, Department of Engineering and Technology,
Old Dominion University, Norfolk, VA 23529.
{Lpott005, Xpalm001}@odu.edu

**Abstract.** Within the field of biocybersecurity, it is important to understand what vulnerabilities may be uncovered in the processing of biologics as well as how they can be safeguarded as they intersect with cyber and cyber-physical systems, as noted by the Peccoud Lab, to ensure not only product and brand integrity, but protect those served. Recent findings have revealed that biological systems can be used to compromise computer systems and vice versa. While regular and sophisticated attacks are still years away, time is of the essence to better understand ways to deepen critique and grasp intersectional vulnerabilities within bioprocessing as processes involved become increasingly digitally accessible. Wargames have been shown to be successful within improving group dynamics in response to anticipated cyber threats, and they can be used towards addressing possible threats within biocybersecurity. Within this paper, we discuss the growing prominence of biocybersecurity, the importance of biocybersecurity to bioprocessing , with respect to domestic and international contexts, and reasons for emphasizing the biological component in the face of explosive growth in biotechnology and thus separating the terms biocybersecurity and cyberbiosecurity. Additionally, a discussion and manual is provided for a simulation towards organizational learning to sense and shore up vulnerabilities that may emerge within an organization's bioprocessing pipeline

## 1   Introduction - Importance of Biocybersecurity to Bioprocessing

Work governing the processing living biological systems and their components falls under bioprocessing. This stretches from the initial research, to the development and manufacturing, and eventually the commercialization of products [1]. Bioprocessing constitutes trillions towards the world economy and continues to grow as novel uses of bio products increase in demand, covering food, fuel, cosmetics, drugs, construction, packaging, and more [1]. As means of biocomputing, bioprocessing, and storage merge and become increasingly accessible and feasible, paired with growing appetites for green technology, the growth of the bioeconomy can be expected to further expand [1-11]. Murch and Richardson point out that entire bioprocessing pipeline is vulnerable to attacks at multiple steps along the way where bioprocessing equipment converges with the internet, which calls to need for additional scrutiny in design and monitoring of organization bioprocessing pipelines [1,6,12-14]. To do so otherwise invites potential disruptions in the world's functioning as the bioeconomy grows [1,4].

Much of the progress in bioprocessing will hinge on increased automation aided by advanced algorithmic processes, increasing engagement with Information Technology, or IT, for short. IT spending, as with that of bioprocessing, has also reached trillions in worldwide spending, with the expectation to continue [15-16]. Paired with open-source methodologies being both profitable and adapted worldwide, economies worldwide are witnessing unprecedented growth in communication and digital technological development [15-23]. Further paired with biological computing and storage, technologies within the bioprocessing pipeline may witness shifts in development that will require new lines of expertise in operation and defense. Nonetheless, basic means of protecting data and processes remain essential [15-23]. As entities engage advanced bioprocessing with aim to run lean operations, it is expected that they will employ networks of connected bioprocessing infrastructure, which will need a great deal of IT expertise for both management and security. For this, it is important that war games be employed by such IT teams to simulate risked posed in their infrastructure.

## 2      Importance of Emphasizing the Bio Component in Biocybersecurity

The primary cause for the development and demarcation of a new field of study is that biological processes are no longer simply the end result of data processing. Biological data is being used as interlocks (retina, fingerprint scanners), to inform decisions (health monitors), and even as the data processing techniques themselves (biocomputing). Thus, in a biocybersecurity context, biological phenomena can act as interlocks, and even as facilitatory steps in a cybersecurity system. Biology, and its manifold aspects introduce numerous targets that can be exploited such as in simple behaviors exhibited by organisms down to the qualities of organic compounds that comprise said organisms. That is, transport of biomolecules within or between organisms can be monitored precisely as well as the targeting of specific genes within an organism [4-5]. These can be examined and targeted on both macro and microscales, requiring creativity, but yielding potentially profitable results. For example, the actor can examine the data from someone's Fitbit to determine their night and daytime activities, for deciphering their target's lifestyle and how to take advantage of that[5, 23-25]. The same can be done for an entire community, in which a company might want to gain additional data on diet or work patterns, to better plan food truck locations. In terms of monitoring and targeting qualities, an insurance company can examine one's genome to glean and set new rates for individuals based off of projected diseases that an individual may develop [26]. Although flawed, it is expected that the tools of analysis and their predictive power will strengthen [27]. In time, as biopro-cessing infrastructure leans on biocomputing, our societies will need to examine such biological systems in an analogous way. In terms of the bioprocessing pipeline, DNA can be used to encode weapons to attack machinery that interfaces with it, but also can be a means of smuggling [13, 28]. As Peccoud and others rightly point out, a naivety in design and protocol can prove disastrous for an organization's enterprise [4,12-13,28]. Thus, teams are encouraged to engage in rigorous examination of the pipeline.

## 3 WarGame Simulation

**Description of and Manual for Wargame Activity for bioprocessing teams:**

The following guide is a fast and simple guide to doing a preliminary biocybersecurity analysis for a bioprocessing center, or any facility that uses biosecurity measures or uses large amounts of biological data:

**Preliminary Step 1** - Selection: Elect or appoint an individual to run the activity. It is recommended that this person has knowledge of Information and cyber security, but any person with active interpersonal skills should do. The gather any individuals interested or with a vested interest in security. Additionally, be sure to screen and update participants on knowledge of biology and biosecurity concepts.

**Preliminary Step 2** - Review: Review the current active security procedures and evaluate participation in them. For instance, investigate how many unsecured devices are in the facility, or how many people leave passwords out or computers unlocked. Now the activity can begin.

**Common Routes of Attack**
- Social Engineering
- Hardware modification
- Bio-Cyber Physical Interfaces
- Supply-Line Corruption
- Phishing

**Common Routes of Defence**
- Mandated Cybersecurity Training
- Segmented Privileges
- Use of Red/Blue Teams
- Ban of personal electronics
- Regular Security Patches

Figure 1: A Review of Attack and Defence Routes to Consider

**Step 1 - Training:** divide the participants into 2 groups. The data defenders and the data hackers. After reviewing the current security protocols as a group, a hacker and defender will pair up. The hacker should attempt to identify a given vulnerability in the current system. The data defender should then attempt to find a way to fix or "patch" that vulnerability. For instance, a facility that has users that frequently use exposed USB SSD devices can be made vulnerable by a malicious actor distributing flash drives with malware on them. Potential defenses include having these devices in the facility, educating users as to the hazards they pose, or disabling inactive USB ports. This section is meant to be an introduction to the key methods used and should only last 5-10 minutes. Then have the participants switch partners, switch roles, or both.

| Data Defender | Data Hacker |
| --- | --- |
| University/Community Bio Lab | Company or Rival Lab looking to gain lab/research/material access or impede research |
| Government Lab / Regulatory Agency | State-Level Hacker from Rival Country looking to gain access to privileged research or product information |
| Company Lab | Rival Company looking to steal/overcome IP and gain a market edge or sabotage company assets |
| Hospital, Prison, | Ransomware Hackers looking to attack patients/equipment to derive ransoms or compromise medical equipment to harm an individual |
| Insurance | Insurance Fraudsters looking to increase costs, skim resources, and or disrupt healthcare access |

Figure 2: Potential Teams: Data Defenders vs Data Hackers

**Step 2 - Group Ideation**: After a few rounds of training, several key ideas or vulnerabilities may become evident. Instead of patching them immediately, at this time, two groups should be formed- again Data Defenders and Data Hackers. This time, they should both be isolated, and allowed to envision defenses against the vulnerabilities or exploits pertaining to vulnerabilities of the facility (respectively). Self selection into the groups is encouraged, as long as both are roughly equal. Allow sufficient time to ideate strategies, and supply materials like adhesive notes, whiteboards, and other common office supplies. The time spent on this stage should be slightly longer than the training stage.

**Step 3 - War Game:** This is the terminal stage of the activity. This section should be recorded in some way in order to analyze weaknesses and potential plans that could ameliorate security vulnerabilities. The activity will take a call-and-response style. Align the groups opposite of each other. The hacker group will announce a plan to exploit a security vulnerability, and the data defender group will have to state a plan to implement mitigation strategies for that plan. This will continue until an impasse is reached or the realistic time limit is reached.

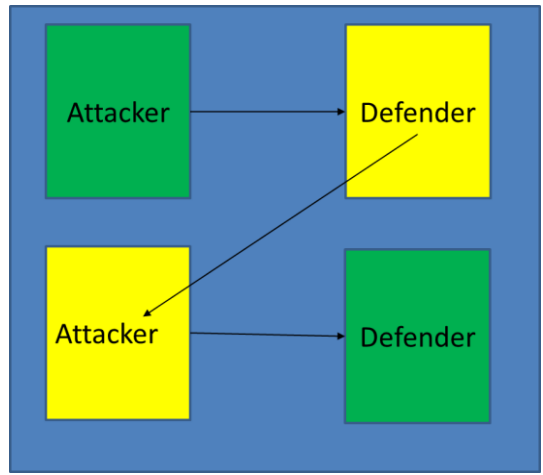

Figure 3: Call and Response Activity Flow

## 4      Discussion

Common exploitations found in trial tuns of this activity have been the following:

1. The inefficiency and potential exploitations of security theater
2. The security implications of underpaid or unpaid workers
3. Miscommunications of conventional security threats
4. Lack of knowledge of novel types of threats
5. Security implications of sub-standard resources

All these combined illustrate that considerable gaps in knowledge exist among potential staff concerning firms that deal in some degree of bioprocessing operations. Regardless of subindustry, it is important that members who take part in such wargames switch sides and remain updated on new trends within biocybersecurity as well as separate developments within cybersecurity and biosecurity, as both fields may have developments that do not yet materially overlap. Further, these war games should be run frequently so as to keep staff in practice and cognizant of dangers that may exist. It is important to note that the style and order of wargaming can be varied to suit the organization's need. That being said, it is suggested and encouraged that teams vary their playstyle and introduce different scenarios. For example, the question of a wargame involving 3 or more opposing groups is worthy of exploration as can be imagined from the interplay of state-level actors, corporate actors, internal actors, and ethical hackers within. As it is increasingly apparent that IT and Bioprocessing possess a shared destiny in the years to come, operations both mental and physical should reflect that for optimal security.

## 5  Importance and Conclusion

The primary importance concerning doing this as a group and as a facility includes the fast-paced nature of development in biology and bioprocessing. Different labs tend to have different requirements, tools, cyber-physical interactions, and workarounds. Lower funded labs may be vulnerable to more conventional means of penetration. Labs with greater amounts of funding could be specifically targeted. Coupled with the fact that almost no research facility will have complete end-to-end supply of anything except the most basic of supplies (for instance ethanol or double distilled water), interactions that could be used and vulnerabilities and exploited for penetration will always exist. Only by the use of representation from all of the invested groups will more complete security coverage be achieved.

**Acknowledgments** We obtained no external funding for this work.